\documentstyle[psfig]{spackap}
\font\s=cmmib10 at 15pt
\begin{opening}
\title {DEVELOPMENT OF A SPECKLE INTERFEROMETER AND THE MEASUREMENT OF FRIED'S
PARAMETER ({\lowercase{\s r$_o$}}) AT THE TELESCOPE SITE}

\author{S. K. \surname{Saha$^1$}} 
\author{G. \surname{Sudheendra$^2$}} 
\author{A. \surname{Umesh Chandra$^2$}} 
\author{V. \surname{Chinnappan$^1$}}

\institute{$^1$Indian Institute of Astrophysics, Bangalore 560034, India 
}
\institute{$^2$Central Manufacturing Technology Institute, Bangalore 560022, India 
}
\date{ }
\runningtitle {DEVELOPMENT OF A SPECKLE INTERFEROMETER }
\runningauthor{{Saha} et. al.} 
\end{opening}

\begin{ao}
S. K. Saha, Indian Institute of Astrophysics, Bangalore 560034, India.\\
e-mail: sks@iiap.ernet.in
\end{ao}

\begin{document}
\begin {abstract} 
A new optical speckle interferometer for use at the 2.34 meter 
Vainu Bappu Telescope (VBT), at Vainu Bappu Observatory (VBO), Kavalur, India,  
has been designed and developed. Provisions have been made for observation both  
at the prime focus (f/3.25), as well as at the Cassegrain focus (f/13) of the  
said telescope. The technical details of this sensitive instrument and the 
design features are described. An interface between the telescope and the 
afore-mentioned interferometer is made based on a concept of eliminating the
formation of eddies due to the hot air entrapment. The performances of this
instrument has been tested both at the laboratory, as well as at the Cassegrain
end of the telescope. It is being used routinely to observe the speckle-grams
of close-binary (separation $<$1 arc second) stars. The size of the Fried's
parameter, r$_{o}$, is also measured.  
\end {abstract} 

\keywords{Interferometer, Speckle Imaging, Image Reconstruction, Seeing}

\section{Introduction}
Speckle interferometric technique (Labeyrie, 1970) has made a major break 
through in observational astronomy by counteracting the effect of atmospheric 
turbulence on the structure of stellar images, allowing measurements of a wide 
range of celestial objects (Barlow et al., 1986, Wood et al., 1987, 
Afanas'jev et al., 1988, McAlister, 1988, Ridgway, 1988, Ebstein et al., 1989, 
Foy, 1992, Saha et al., 1997a, Saha and Venkatakrishnan, 1997). The 
principle of this technique is described in detail in the literature (Dainty,
1975). A new instrument based on the above technique has been developed to 
operate at both the foci, prime (f/3.25), as well as Cassegrain (f/13) of the 
2.34 meter Vainu Bappu Telescope (VBT) at Vainu Bappu Observatory (VBO), 
Kavalur, India. Since the afore-mentioned interferometer is a very sensitive 
instrument, any inaccuracies will lead to erroneous conclusion of the 
observations. Therefore, emphasis was given on the accuracies of the mechanical 
mounts and housing so as to ensure the optical path precisely. Most of the 
critical mounts were machined out of a solid piece. Care is taken in the design 
analysis and manufacturing to get required dimensional and geometrical 
accuracies. The salient features of important elements are described in this 
paper. Finally, we present the size of r$_{o}$, measured from 
the speckle-grams of $\alpha$-Andromeda obtained with this interferometer at 
the VBT, Kavalur. 

\section{Design Features}
\subsection{Optical Design}
Figure 1 shows the optical design of the speckle camera system. Provisions have 
been made in the design to observe at both the foci of the 2.34 VBT. The wave 
front falls on the focal plane of an optical flat made of low expansion glass 
with a high precision hole of aperture ($\sim$350 $\mu$), at an angle of 
15$^{o}$ on its surface (Saha et al., 1997). The image of the object passes on 
to the microscope objective through this aperture, which slows down the image 
scale of this telescope to f/130. A narrow band filter is placed before 
the detector, to avoid the chromatic blurring. The surrounding star field of 
diameter $\phi$ 10 mm, gets reflected from the optical flat on to a plane 
mirror and is re-imaged on an intensified CCD, henceforth ICCD, (Chinnappan
et al., 1991). 

\begin{figure}[h]
\centerline{\psfig{figure=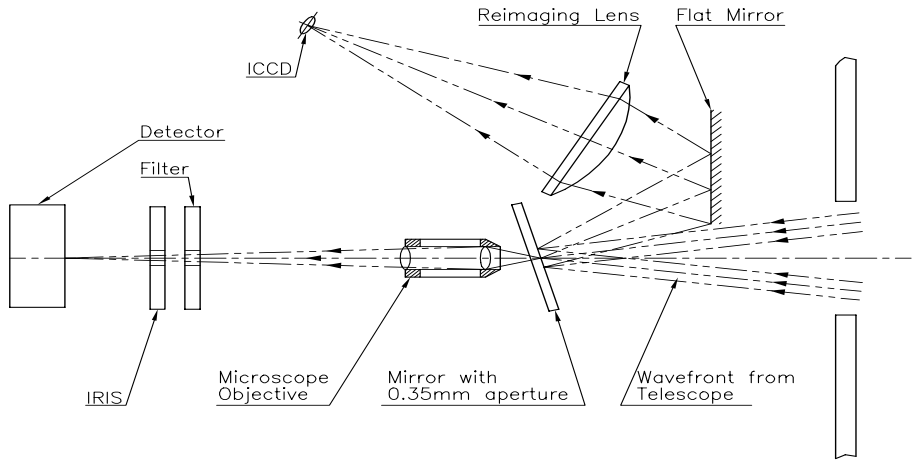,height=6cm,width=12cm}}
\caption{
Optical design of the Speckle Interferometer.
}
\end{figure}

\subsection{Mechanical Design Requirements} 
The instrument has been designed with the help of the Computer Aided Design 
and analysis techniques and manufactured accurately on precision machine tools.
The instrument poses the following requirements. The instrument should primarily
be light in weight besides being rigid when subjected to bending loads arising 
out of multiple orientations of the telescope. The deflection of the instrument 
should be minimum under its own weight, as well as when it is fitted with all 
accessories required for observation. Provisions for fine focusing of the image 
is to be incorporated for better resolution and clarity in observation. Besides 
these, it has to be provided with linear motion systems with no rotation of the 
optical elements and rigid locking of the lens position. The instrument should 
be aloof from temperature effects. The instrument is made of Martensitic 
Stainless Steel (SS410), a material with low co-efficient of linear expansion 
(about half that of normal steel).  

\subsubsection{Design Analysis}
Since the optical enclosure of the interferometer has to support  
various mounts and the detector without any deflection in various orientations 
of the telescope, before arriving at the final configuration and sizes, finite
element method (FEM) was used to analyze (Zienkiewicz, 1967) as well as to have 
prior information about the deflections, stress, flexure etc. of the enclosure. 
For structural analysis, Pro-mechanica software has 
been used for optimizing the structural members. The analysis
shows that the instrument can hold any detector or camera of 20 Kgs. weight, 
kept at a distance of 270mm away from the mounting flange of the interferometer 
undergoes a maximum deflection of 1.65$\mu$. The model was analyzed for strength
and deflection for a distributed load of 20 Kgs. over the span of the 
instrument. Deflection of only 0.7$\mu$ was observed. Figure 2 shows the FEM 
model of the speckle interferometer. It depicts the analysis carried out for two
kinds of loads in the two windows. Left window shows the deformation pattern
for load 1 and right window shows deformation pattern for load 2. The 
grey-scale blocks in the picture indicate the extent of 
deformations at individual places in the model of the interferometer.  

\begin{figure}[h]
\centerline{\psfig{figure=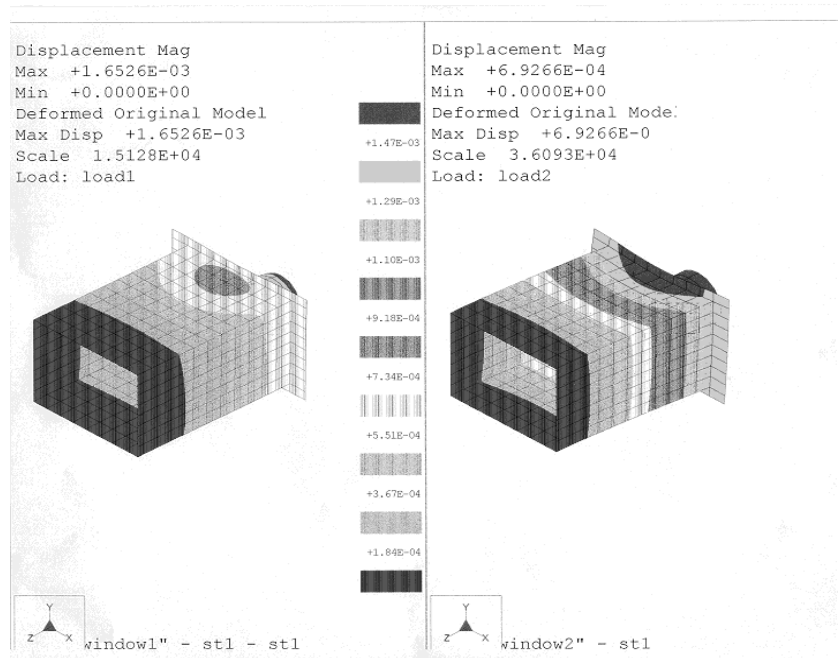,height=9cm,width=11cm}}
\caption{
Finite Element Model of Speckle Interferometer.
}
\end{figure}

\subsection{Elements of the Interferometer}
An overall view of the interferometer is shown in Figure 3. It can be seen that 
the instrument basically consists of an optical enclosure, which houses the 
following assemblies. (i) Microscope assembly, (ii) Filter assembly, 
(iii) Detector mount, (iv) Guiding system.  

\begin{figure}[h]
\centerline{\psfig{figure=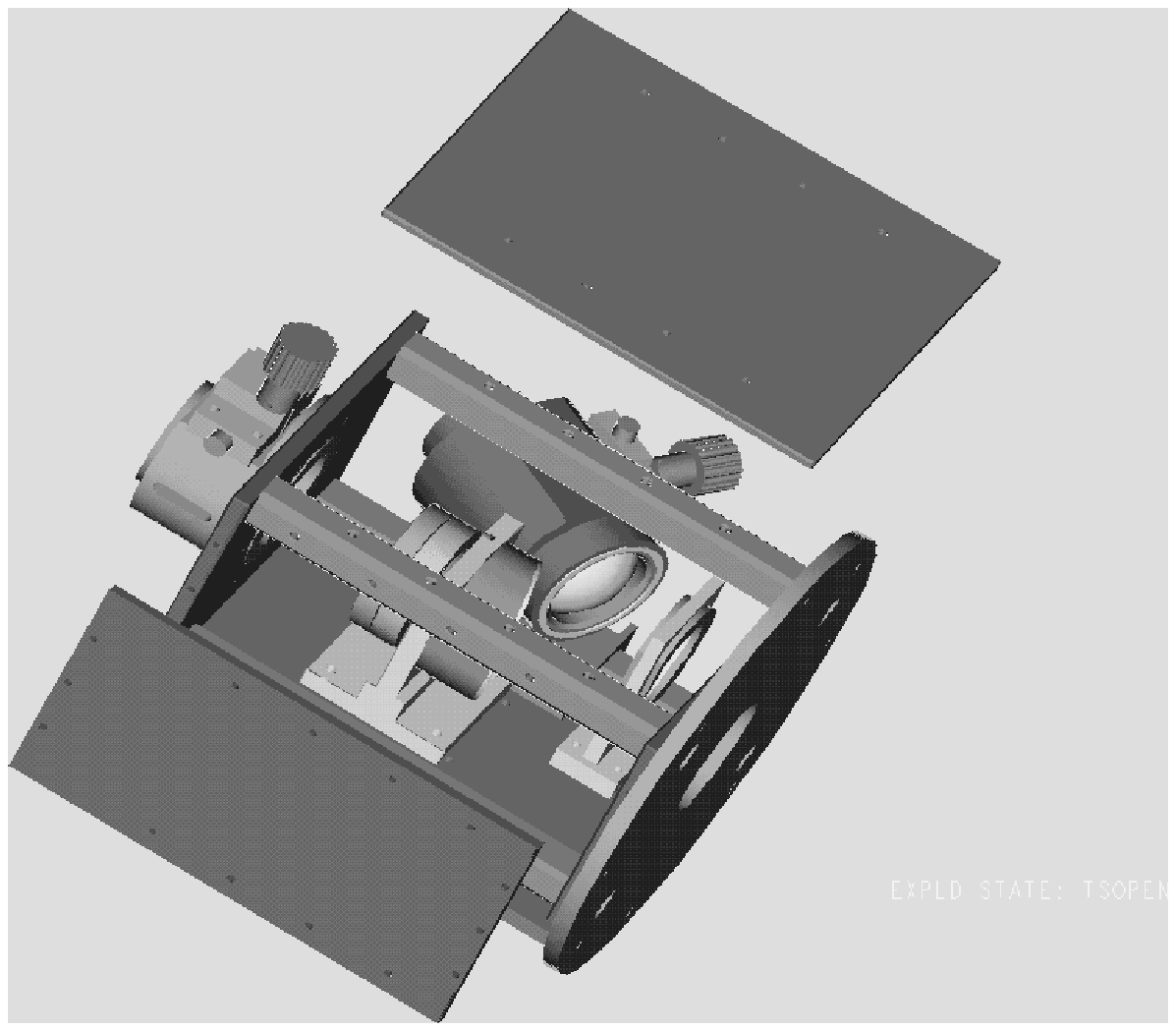,height=9cm,width=11cm}}
\caption{
Overview of the Speckle Interferometer.
}
\end{figure}

\subsubsection{Optical Enclosure}
The instrument has been conceived as a box made of two end plates of thickness 
8mm and joined by means of struts of section 22mm square. The struts have 
been machined from cylinders of $\phi$ 25mm diameter and required length. The 
struts are provided with spigots of $\phi$ 18mm diameter at ends for locating 
the end plates. The diameters of $\phi$ 25mm and $\phi$ 18mm at both ends are 
ground between centers to ensure concentricity of the spigots and the main 
cylinder. The end plates are provided with four holes to receive the spigots and
are machined together in one setting on Wire-cut Electro Discharge Machining 
(EDM) to assure the size and the required center distance accuracies. The end 
plates, when locked with the struts, form a box structure of light weight with 
the required strength to house the designed mounts with minimum deflection. A 
bottom plate has been clamped on to two of the struts and this forms the 
platform on which the various mounts of the microscope can be mounted. Once the 
reference is established parallel to the path of the wave front, the mounts can 
be manufactured in such a way that the lens mounting holes are at an accurate 
distance from their bases. In case of any error, the base of the mounts can 
be ground to get the required center height. Thus the plane in which the light 
travels through the microscope and the mirrors is maintained accurately. One 
of the sides has been provided with side plates shaped in such a way as to 
reduce overall volume occupied by the enclosure while ensuring no light 
leakage. The optical enclosure has been blackened to ensure that no stray 
light is received by the optical elements. A special process has been used to 
blacken the stainless steel material.

\subsubsection{Microscope Assembly}
The microscope mount assembly is shown in figure 4. The main body of the 
assembly has a base and a vertical face in which a bore has been made to receive
the focal plane mirror and the microscope. The axis of the bore is maintained 
parallel to the base and its distance is maintained accurately like all the 
other mounts. The focal plane mirror mount is mounted at an angle of 
15$^{o}$ with the axis of the aperture (350$\mu$) as mentioned earlier (see
section 2.1). After mounting the focal plane mirror, the aperture comes in the 
axis of the microscope mount. The image of the object is passed on to the 
microscope objective through this aperture. In the design, fine focusing
arrangement has been provided to get optimum focus of the image. Accurate 
aligning, guiding and locating have been ensured to get an accurate straight 
line motion of the objective along the direction of optical axis. A rigid 
locking system has been provided taking into consideration that the image 
should not be disturbed either during or after locking. Care has also been taken
to ensure that the axes of the 15$^{o}$ bore and the optical axes intersect at a
predetermined point. A mirror cell has been conceived and designed to facilitate
the focal point flat elliptical mirror ensuring its safety into consideration.  

Further development of this incorporates a nano-adjusting mechanism which helps
in ultra fine focusing of the microscope objective. Flexure mechanism is being 
used to achieve the nanometric motion of the same.

\begin{figure}[h]
\centerline{\psfig{figure=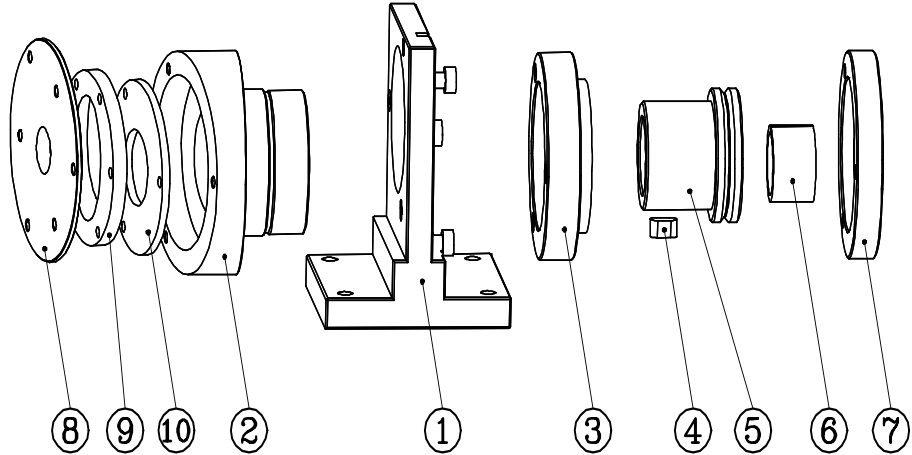,height=6cm,width=12cm}}
\caption {
Microscope Objective Assembly. The numbers in the figure 
depicts as: 1. Mount, 2. Mirror Housing, 3, 4, 5, 6 \& 7. Objective Moving 
Mechanism and 8, 9 \& 10. Mirror Cell.
}
\end{figure}

\subsubsection{Detector Mount}
The high speed electronic detectors are being used to record the speckle-grams
of faint point source. This may be in the form of either detecting the photon 
events per frames $-$ up to frame rate of 50 Hz (Blazit et al., 1977, Blazit, 
1986), or detecting individual photon events with time resolution of a few
$\mu$sec (Papaliolios et al, 1985, Durand et al., 1987, Graves et al., 1993). In
this design too, we made an arrangement for fine focusing the speckles by moving
the detector back and forth precisely in line with the optical axis. Locking 
arrangement has been provided to avoid effects of backlash. In addition, we
made an arrangement for the micrometric x$-$y movement of the detector which 
would ensure its position precisely. A good quality manually adjusted iris  
kept in front of the detector, eliminates the extraneous light. Provision for
mounting different narrow band filters for observation have also been made.

\subsubsection{Guiding System}
The surrounding star field of a few arc min gets reflected from the elliptical 
mirror and falls on to the flat mirror at an angle of 30$^{o}$ with respect to 
the optical axis. The mount of the flat mirror has been relieved suitably to 
ensure that the image reaches from the elliptical mirror without any 
obstructions and image is fully received by the flat mirror.
The re-imaging lens mount is designed with provisions for lens mounting, 
focusing and a rigid locking mechanism. The lens collects the reflected rays 
from the afore-mentioned optical flat and is re-imaged on to an ICCD so as to 
enable one to guide remotely from the Console room of the telescope.

\subsubsection{Spacer Assembly}
It is found that the best focus at the Cassegrain end of the 2.34 meter VBT
is at a distance $\sim$570 mm away from its mounting flange. 
In the design of this interferometer, we have kept the focal point,
the aperture (350$\mu$) on the elliptical mirror, as described in section 2.1,
110mm away from the surface of the mounting flange at the Prime focus of this 
telescope. Therefore, it is essential to design a spacer of $\sim$450mm size, 
so as to enable us to observe at the Cassegrain end of the VBT.
The design of the spacer has to satisfy the following requirements : 
light in weight, high rigidity and low deflection at various orientations of 
the telescope. When mounted, both the mounting diameters should be perfectly 
concentric to each other and it should not allow the formation of eddies that 
may be produced due to hot air entrapment. In order to satisfy the above 
requirements, spacer is designed with two plates separated by six pillars of 
equal height. The pillar rods have concentric spigots at both ends. These 
spigots are located in holes provided on pitch circle diameter concentric 
with respect to the reference diameters. The pillars are rigidly bolted on to 
the plates at both ends. This design ensures a rigid, light weight spacer 
which allows free movement of air and has concentric diameter plates on both 
ends. Figure 5 depicts the design of the spacer. 

\begin{figure}[h]
\centerline{\psfig{figure=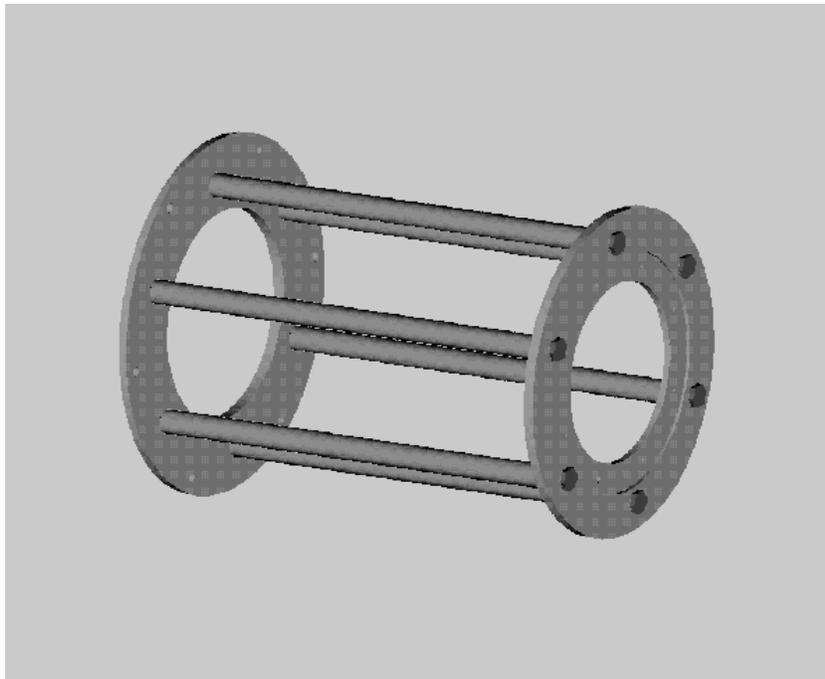,height=9cm,width=11cm}}
\caption{
Spacer Assembly.
}
\end{figure}

\section{Observations at VBT} 
Observations of several close binary systems (separation $<$1 arc sec), and of 
other stars, using this newly built interferometer have been successfully 
carried out on 29/30 November, 1996, at the Cassegrain focus of the VBT through 
a 5nm filter centered on H$\alpha$ using uncooled ICCD. The image scale at the 
afore-mentioned focus of this telescope is magnified to 0.67 arcsecond per mm. 
The CCD gives video signal as output. The images were acquired at an exposure 
times of 20 ms using a frame grabber card DT-2861 manufactured by Data 
Translation{$^T$$^M$} and stored on to the hard disk of PC486 computer. It can 
store upto 64 frames in a few seconds time. The interface between the 
intensifier screen and the CCD chip is a fibre-optic bundle which reduces the 
image size by a factor of 1.69. The frame grabber card digitises the video 
signal and resamples the pixels of each row (385 CCD columns to 512 digitised 
samples), by introducing a net reduction in the row direction of a factor of 
1.27. These images were analysed on a Pentium PC.

\section{Measurement of r$_{o}$ }
The resolution $\theta$ of a large telescope, limited by the atmospheric
turbulence, as defined by the Strehl criterion is 

\begin{equation}
{\theta = {\frac {4}{\phi}} {\frac {\lambda}{r_{o}}}}
\end{equation}

where $\lambda$ is the wavelength of observation and r$_{o}$ is the Fried's
parameter. 

Hence, r$_{o}$ is directly related to the seeing at any place. The conventional
method of measuring seeing from the star image is to measure the full width half
maximum (FWHM) of a long exposure stellar image at zenith:

\begin{equation}
{FWHM = {0.98 {\frac {\lambda}{r_{o}}}}}
\end{equation}

There are different methods of measuring the seeing at any given place.
Through a large telescope, the size of the seeing disk is often estimated by
comparing it to the known angular separation of a double star. The seeing value
can be estimated by star trails too. The other qualitative method is to measure
r$_{o}$ from the short exposure images using speckle interferometric technique.
The speckle life time is an important atmospheric parameter since it describes
the longest possible exposure time for recording speckle interferograms. It
is an important parameter too for testing the atmospheric condition at 
existing and incoming astronomical sites (Vernin et al., 1991). In 
this technique, the area of the telescope aperture divided by the estimated
number of speckles gives the wavefront coherence area $\sigma$, from which
r$_{o}$ can be found by using relation, 

\begin{equation}
{\sigma = {0.342 {\left( \frac {r_{o}} {\lambda}\right) }^{2}}}
\end{equation}

Von der L$\ddot{u}$he (1984) suggested that the squared modulus of the average 
observed Fourier transform should be divided by the averaged observed power 
spectrum. This expression depends only on the seeing conditions that prevails 
during the period covered by the time series and on the signal-to-noise ratio. 
The method may be useful for the observations of extended object where reference 
is not available.

\begin{figure}[h]
\centerline{\psfig{figure=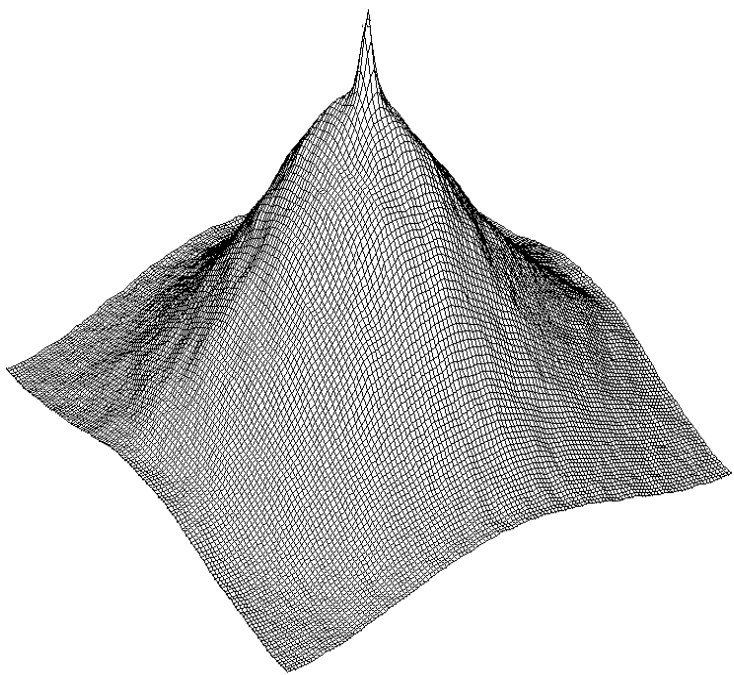,height=9cm,width=11cm}}
\caption{
Autocorrelation of seeing disk, as well as of speckle component.
}
\end{figure}

If many short exposure images and their autocorrelations are summed, the summed 
images have the shape of seeing disk, while the summed autocorrelations contain 
autocorrelation of the seeing disk together with the autocorrelation of the 
mean speckle cell (Wood, 1985). It is width of the speckle component of the 
autocorrelation that provides information on the size of the object being 
observed.

We have used the speckle interferometric technique to measure the Fried's 
parameter at 2.34 meter VBT, VBO, Kavalur. 10 continuous speckle-grams of 
$\alpha$-Andromeda, were processed. Figure 6 depicts the autocorrelation
of the seeing disk together with the autocorrelation of the mean speckle cell.
The size of the r$_{o}$ is found to be the 11.44cm at the FWHM.  

\section{Discussions and Conclusions }
Optimized instruments and meticulous observing procedures are part of the
important ground work for addressing the basic astrophysical problems. The
new interferometer would enable one to observe many interesting objects and
to map their high resolution features, taking the positional advantage of the
site, at Kavalur. The latitude of this observatory (12.58$^{o}$) gives us 
access to almost 70$^{o}$ south of the celestial equator, therefore, most of the
observational results beyond 30$^{o}$ south of zenith at high latitude stations 
obtained earlier can be confirmed. Before arriving at the final concept of the
design of this instrument, several experiments were carried out at the various
telescopes (Saha et al., 1987, Chinnappan et al., 1991), as well as at the 
laboratory (Saha et al., 1988). Emphasis was given on the design analysis using
the modern FEM technique while designing the instrument to obtain required 
dimensional and geometrical accuracies of the mechanical mounts and house so as 
to avoid erroneous conclusion of the observations.

The quality of the image degrades due to the following reasons: (i) variations
of air mass X ($\sim$ 1/cosZ) or of its time average between the object 
and the reference, (ii) seeing differences between the modulation transfer 
function (MTF) for the object and its estimation from the reference,
(iii) deformation of mirrors or to a slight misalignment while changing
its pointing direction, (iv) bad focusing, (v) thermal effect from
the telescope etc. (Foy, 1988). Measurement of r$_{o}$ is of paramount 
importance to estimate the seeing at any astronomical site. Systematic studies 
of this parameter would enable to understand the various causes of the local 
seeing, for example, thermal inhomogeneities associated with the building, 
aberrations in the design, manufacture and alignment of the optical train etc. 
Significant improvement of the seeing was noticed during the observations of the
speckles of close binaries, by introducing the afore-mentioned spacer, that 
does not allow the formation of eddies produced by the hot air
entrapment, as an interface between the telescope and the interferometer 
compare to earlier observations (Saha and Chinnappan, 1998). 

\acknowledgements The authors are grateful to Prof. J C Bhattacharyya,
for the constant encouragement during the progress of the work. The personnel
of the mechanical division of IIA, in particular Messrs. B R Madhava Rao,
R M Paulraj, K Sagayanathan and A Selvaraj provided excellent support during
execution of the work. The help rendered by Mr. J R K Murthy and Y V Ramana 
Murthy of CMTI for the computer analysis of the design and by Dr. Indira
Rajagopal of National Aerospace Laboratory, Bangalore for the black chrome
plating are also gratefully acknowledged.


\end{document}